\documentclass[12pt]{article}
\usepackage{subeqn}
\usepackage{epsfig}
\newcommand{\be}{\begin{equation}}
\newcommand{\ee}{\end{equation}}

\newcommand{\no}{\noindent}
\newcommand{\ce}{\begin{center}}
\newcommand{\nc}{\end{center}}

\makeatletter
\@addtoreset{equation}{section}
\makeatother

\def\sqr#1#2{{\vcenter{\vbox{\hrule height.#2pt
 \hbox{\vrule width.#2pt height#1pt \kern#1pt
 \vrule width.#2pt} \hrule height.#2pt}}}}

\def\operp{\hbox{${\kern+.25em{\bigcirc}
\kern-.85em\bot\kern+.85em\kern-.25em}$}}

\def\lsim{\;\raise0.3ex\hbox{$<$\kern-0.75em\raise-1.1ex\hbox{$\sim$}}\;}
\def\gsim{\;\raise0.3ex\hbox{$>$\kern-0.75em\raise-1.1ex\hbox{$\sim$}}\;}
\def\no{\noindent}

\def\ce{\centerline}
\def\ve{\vfill\eject}
\def\rdots{\mathinner{\mkern1mu\raise1pt\vbox{\kern7pt\hbox{.}}\mkern2mu
 \raise4pt\hbox{.}\mkern2mu\raise7pt\hbox{.}\mkern1mu}}

\def\e e{$e^+ e^-$ }

\begin{document}

\ce{\bf On the Deformation Parameter in SLq(2) Models of the Elementary Particles}
\vskip.3cm

\ce{\it Robert J. Finkelstein}
\vskip.3cm

\ce{Department of Physics and Astronomy}
\ce{University of California, Los Angeles, CA 90095-1547}

\vskip1.0cm

\no {\bf Abstract.}   When the fundamental invariant of $SLq(2)$ is expressed as $\varepsilon_q = \left(\matrix{ 0 & \alpha_2 \cr -\alpha_1 & 0} \right)$, then the deformation parameter, $q$, defining the knot algebra is $q = \frac{\alpha_1}{\alpha_2}$.  We consider models in which the elementary particles carry more than one kind of charge with running coupling constants, $\alpha_1$ and $\alpha_2$, having different energy dependence and belonging to different gauge groups.  Let these coupling constants be normalized to agree with experiment at hadronic energies and written as $\alpha_1 = \frac{e}{\sqrt{\hbar c}}$ and $\alpha_2 = \frac{g}{\sqrt{\hbar c}}$.  Then $q = \frac{e}{g}$.  If $e$ is an electroweak coupling and $g$ is a gluon coupling, $q$ will increase with energy.  In previous discussions of $SLq(2)$ it has been assumed that $\varepsilon_q^{2} = -1$.  If this condition is maintained, then $eg = \hbar c$.  If the elementary particle is like a Schwinger dyon and therefore the source of magnetic as well as electric charge, $eg = \hbar c$ is the Dirac condition for magnetic charge.

\ve

\section{Introduction}

\hspace{10mm} In the standard electroweak model one characterizes the elementary particles by their momentum, spin, isotopic spin, and hypercharge; these are properties that may be related to different realizations of $SL(2C)$, $SU(2)$,and $SU(2)$x$U(1)$ symmetry.  The "knot" model adds to the standard model of the elementary particles, a topological symmetry$^{(1)(2)(3)(4)}$, as expressed by the knot algebra, $SLq(2)$.  This additional symmetry permits a uniform description of the four classes of elementary fermions and has two major consequences: the first describes modified interactions between elementary fermions, where the standard matrix elements are multiplied by $SLq(2)$ form factors$^{(2)(3)(4)}$, and the second describes a preonic structure$^{(5)(6)(7)}$ of the elementary particles of the standard theory.  The model is empirically based on electroweak physics.  Here we discuss possible physical interpretations of the free parameter $q$ as it appears in the $SLq(2)$ deformation of the standard model but also as it might appear in a wider class of models.

\section{The Fundamental Representation of SLq(2)}

\hspace{10mm}  The fundamental representation, $T$, of the $SLq(2)$ algebra may be defined by
\be
T^t\varepsilon_qT = T\varepsilon_qT^t = \varepsilon_q
\ee

where

\be
\varepsilon_q=\left(\matrix{0 & q_1^{\frac{1}{2}} \cr -q^{\frac{1}{2}} & 0} \right)
\ee

and where we have previously taken

\be
q_1=q^{-1}
\ee

Then

\be
\varepsilon_q^2 = -1
\ee

\hspace{10mm}  In the mathematical literature $q$ appears simply as a deformation parameter or as the independent variable in the Jones polynomial that labels a knot.  In the applications of the $SLq(2)$ (knot) model that we have made to electroweak physics, $q$ is considered a dimensionless parameter to be determined by the data.  It is a measure of the deformation imposed on the standard model by the knotting of the elementary fermions.  It appears to be energy dependent and resembles a running coupling constant that turns out to be in the neighborhood of unity for these applications.  In this note we try to refine and generalize the possible physical meaning of $q$.

\hspace{10mm} Let us first reformulate $SLq(2)$ by dropping (2.3) and (2.4), which depend on the single parameter, $q$, and replacing (2.2) with

\be
\varepsilon_\alpha = \left( \matrix{ 0 & \alpha_2 \cr -\alpha_1 & 0} \right)
\ee

\noindent depending on the two parameters $\alpha_1$ and $\alpha_2$.

\hspace{10mm} Now denote the fundamental representation of $SLq(2)$ by

\be
T= \left( \matrix{a & b \cr c & d} \right)
\ee

\noindent Then by (2.1), (2.5), and (2.6)

\[
\begin{array}{llllr}
ab = \alpha ba & bd =\alpha db & ad - \alpha bc =1& bc=cb \\
ac=\alpha ca & cd=\alpha dc & da-\alpha^{-1} cb=1 & & \hspace{1.25in}\mbox{(A)}
\end{array}
\]

\noindent
where

\be
\alpha= \frac{\alpha_1}{\alpha_2}
\ee

\noindent
Then $\alpha$ replaces the $q$ that appears in our earlier description of the knot algebra

\hspace{10mm} In the knot electroweak model it is natural to relate $\alpha_1$ and $\alpha_2$ to the writhe and rotation charges, $Q_w$ and $Q_r$, respectively, that are in turn related to the standard quantum numbers, ($t_3, t_0$) as follows$^{(7)(9)}$:

\be
Q_w = 3k_w t_3
\ee
\be
Q_r = 3k_r t_0
\ee

where $k_w$ and $k_t$ have the dimensions of electric charge and are empirically determined.

\hspace{10mm} By (2.8) and (2.9)

\be
Q_w + Q_r = 3(k_w t_3 + k_r t_0)
\ee

\hspace{10mm} If $k_w = k_r = \frac{e}{3}$, then

\be
Q_w + Q_r = e(t_3 + t_0)
\ee

\noindent is the total charge as expressed in the standard model.

\noindent Now set

\be
\varepsilon_q = \left( \matrix{ 0 & k_w \cr -k_r & 0} \right)
\ee

\noindent
Then
\be
\alpha = \frac{k_w}{k_r} = \frac{Q_w t_0}{Q_r t_3}
\ee

\noindent
Empirical data on lepton-neutrino and quark-quark interactions (form factors) require $\alpha$ to be near unity$^{(2)(9)}$ but the relative masses of the three generations of fermions imply that $\alpha$ is greater than unity$^{(2)}$.  We attribute this variation in $\alpha$ to an energy dependence that is different for $k_w$ and $k_r$.  With this interpretation the data require

\be
\alpha \left( E_H \right) \cong 1
\ee

\noindent where $E_H$ is a hadronic energy in the neighborhood of $E_Z$, the mass of the $Z$, but

\be
\alpha \left( E_p \right) > 1
\ee

\noindent where $E_p$ indicates a renormalization scale appropriate for the determination of particle mass and much greater than the mass of the $Z$.

\hspace{10mm}  Note that if a dimensional form of (2.4) is reinstated as a necessary condition for defining the model, one may write

\be
k_w k_r = \frac{e^2}{9}
\ee

\noindent so that $k_w$ and $k_r$ vary inversely as the energy is increased, and where (2.14) holds, one also has

\be
k_w = k_r = \frac{e}{3}
\ee

\noindent as in (2.11), correct for the standard model.

\section{The Weinberg-Salam Parameters}

\hspace{10mm} In the Weinberg-Salam model

\be
e = g \sin \theta_w
\ee
\be
e = g' \cos \theta_w
\ee
\be
\frac{g'}{g} = \tan \theta_w
\ee

\noindent where $g$ and $g'$ are coupling constants of the $SU(2)$ and $U(1)$ groups, respectively, and where $\theta_w$ is the Weinberg angle.

\hspace{10mm} If we set
\be
\varepsilon_q = \left( \matrix{ 0 & g\sin\theta_w \cr -g'\cos\theta_w & 0 } \right)
\ee

\noindent then

\be
\alpha=\frac{g'}{g}\cot\theta_w
\ee

\noindent and by (3.3) and (3.5)

\be
\alpha = 1
\ee

\hspace{10mm}  The equations of the Weinberg-Salam model hold at hadronic energies near the mass of the $Z$.  According to renormalization theory, g$^\prime$ increases and g decreases with increasing energy.  Assuming that $\theta_w$ is fixed (at $\sin \theta_w$ = 0.218), $\alpha$ also increases with increasing energy and at some high energy, $E_p$, the curves $g \left( E \right)$ and $g' \left( E \right)$ will cross and

\be
\alpha \left( E_p \right) = \cot \theta_w = 4.47
\ee

\noindent This model therefore predicts weak knotting at hadronic energies and allows stronger knotting at higher energy.

\hspace{10mm}  Both of the physical interpretations of $\alpha$ that we have considered depend on the fact that there are two contributions to the electric charge, which we have described as coming from either the $g$ and $g'$ fields of the standard model or as coming from the $Q_w$ and $Q_r$ charges of the knot model.  In both descriptions, $\alpha$ needs to be adjusted to fit the experimental input at hadronic energies near the mass of the $Z$.  In knot parameters we require that at the mass of the $Z$

\begin{center}
$k_w = k_r = \frac{e}{3}$ and $\alpha = \frac{k_w}{k_r} =1$
\end{center}

\noindent and in the Weinberg-Salam parameters

\begin{center}
$\alpha = \frac{g'}{g} \cot \theta_w =1$
\end{center}

\noindent at the same energy.

\hspace{10mm}  These relations hold near the mass of the $Z$ but the knot deformation as measured by $\alpha$ will increase as the energy is increased.

\section{More General Models}

\hspace{10mm}  One may ask whether knotting is possible more generally whenever there are two fields of different symmetry having their sources on the same particle.  Let us then consider two fields of different symmetry with running coupling constants, $\alpha_1 \left(E \right)$ and $\alpha_2 \left( E \right)$, with a different dependence on the energy so that $\alpha = \frac{\alpha_2}{\alpha_1}$ is also energy dependent.

\noindent Then define

\be
\varepsilon_\alpha = \left( \matrix{ 0 & \alpha_2 \left(E \right) \cr -\alpha_1 \left(E\right) & 0} \right)
\ee

\noindent After $\alpha_1 \left( E \right)$ and $\alpha_2 \left( E\right)$ are adjusted to agree with experiment at hadronic energies, we shall write the adjusted $\alpha_1$ and $\alpha_2$ as $\frac{e}{\sqrt{\hbar c}}$ and $\frac{g}{\sqrt{\hbar c}}$.

\noindent Then

\be
\varepsilon_\alpha = \left( \matrix{0 & \frac{g \left(E \right)}{\sqrt{\hbar c}} \cr -\frac{e\left(E \right)}{\sqrt{\hbar c}} & 0} \right)
\ee

\noindent and

\be
\alpha = \frac{e}{g}
\ee

\noindent  Then, if $e$ and $g$ are electroweak and gluon coupling constants, respectively, $g$ will decrease and $e$ will increase, and so the knotting, as measured by $\alpha = \frac{e}{g}$, will increase with energy.

\hspace{10mm}  In the Schwinger dyon model$^{(8)}$, where the elementary particle carries both electric and magnetic charge, $e$ and $g$ are electric and magnetic charge, and the energy dependence of $e \left(E \right)$ and $g \left(E \right)$ will depend on the dyon dynamics.

\hspace{10mm}  We again note that if (2.4) is an essential condition defining the model, then

\be
eg = \hbar c
\ee

\noindent for both the electroweak-gluon model and for the dyon model.  In the dyon case, (4.4) is the Dirac condition connecting electric and magnetic charge.

\hspace{10mm}  If $g$ is either gluon or magnetic charge, attraction between particles of opposite charge must be very strong at hadronic energies to be consistent with the familiar absence of isolated quarks and magnetic monopoles.  If $e$ and $g$ move according to the basic scenario, however, $g$ will decrease and $e$ will increase with increasing energy so that at some point we could have

\be
e\left(E_p \right) = g\left(E_p \right)
\ee

\noindent where $E_p$ may be much higher than $E_Z$ and is possibly a preonic energy.  At this point one could have

\be
\matrix{\alpha = 1 \cr e =g = \sqrt{\hbar c} }
\ee

\section{The $\alpha$-Dependence of the Knot Model}

\hspace{10mm}  The knot modified theory is constructed by adjoining $\mathcal{D}^j_{mm'}$ to the field operator of the standard model, similar to the way the spin is introduced by attaching a spin state.  If the standard field operator is $\psi (t, t_3, t_0)$, the knot modified field operator is $\psi (t,t_3,t_0) \mathcal{D}^j_{mm'}\left( \alpha \right)$ where the $(t,t_3,t_0)$ are isotopic spin quantum numbers and the $(j, m,m')$ are related to the $(t, t_3,t_0)$ by

\be
(j,m,m')=3(t,-t_3,-t_0)
\ee

\noindent The $(j,m,m')$ are also restricted by the following explicit connection with the spectrum of a classical knot

\be
(j,m,m')=\frac{1}{2}(N,w,r+1)
\ee

\noindent where $(N,w,r)$ are the numbers of crossings, the writhe, and the rotation of the corresponding classical knot.

\hspace{10mm}  By (5.1) the elementary fermions with $t=\frac{1}{2}$ lie in the $j=\frac{3}{2}$ representation of $SLq(2)$ and by  (5.2) they correspond to the $N=3$ classical trefoil.  Then the most elementary particles $(t=\frac{1}{2})$ correspond to the simplest knots $(N=3)$.

\hspace{10mm}  By (5.1) the $j=\frac{1}{2}$ and $j=1$ representations of $SLq(2)$ provide an extension of the isotopic spin group and define presently unknown particles.  We refer to the elements of the fundamental representation $\mathcal{D}^{\frac{1}{2}}_{mm'}$ as preons and elements of the adjoint representation $\mathcal{D}^1_{mm'}$ as bosonic preons.  

\hspace{10mm}  The $2j+1$-dimensional representation of the $SLq(2)$ algebra may be written as$^{(7)}$

\be
\mathcal{D}^j_{mm'} \left( \alpha \right) = \sum_{\matrix{0 \le s \le n_+ \cr 0 \le t \le n_-}} \mathcal{A}^j_{mm'} \left( \alpha, s, t\right) \delta \left( s+t,n_+' \right) a^s b^{n_+ - s} c^t d^{n_- -t}
\ee

\noindent where

\be
\alpha = \frac{\alpha_1}{\alpha_2}
\ee

\be
\matrix{n_{\pm} = j \pm m \cr n'_{\pm} = j \pm m'}
\ee

\noindent When $\alpha = 1$, the $\mathcal{D}^j_{mm'}$ are irreducible representations of $SU(2)$.  The coefficients $\mathcal{A}^j_{mm'}$ as well as the knot algebra (A) depend on $\alpha$.  According to (5.3) the general representation $\mathcal{D}^j_{mm'}$ is expanded in elements of the fundamental representations (a,b,c,d).

\hspace{10mm} The form factors depend on $\alpha$ through $\mathcal{D}^j_{mm'} \left( \alpha \right)$ in the following matrix elements$^{(2)(7)}$:

\be
\langle n_3 \left | \right. \bar{\mathcal{D}}^{j_3}_{m_3m_3'}\left( \alpha \right) \mathcal{D}^{j_2}_{m_2m_2'}\left( \alpha \right) \mathcal{D}^{j_1}_{m_1m_1'} \left( \alpha \right) \left | \right. n_1 \rangle
\ee

\noindent where the $\left | \right. n \rangle$ are eigenstates of the commuting $b$ and $c$.  By comparing (5.6) with experimental data, and in particular with the Cabbibo-Kobayashi-Maskawa matrix, one finds that $\alpha \cong 1$ and $\beta \cong 1$ as well, where $\beta$ is the eigenvalue of $b$ on the ground state.  The relative masses of the three members of each fermion family depend on similar but diagonal matrix elements$^{(1)(2)}$

\be
\langle n \left | \right. \bar{\mathcal{D}}^{\frac{3}{2}}_{mm'}\left( \alpha \right) \mathcal{D}^{\frac{3}{2}}_{mm'}\left( \alpha \right) \left | \right. n \rangle
\ee

\noindent where $m \left ( = -3t_3 \right)$ and $m' \left( = -3t_0 \right)$ label the family and $n$ labels the member of the family.  These matrix elements are polynomials in $\alpha$.  By comparing (5.7) with the observed masses of the elementary fermions, one finds that $\alpha >1$ and $\beta >1$.

\hspace{10mm}  One arrives at the preon model by interpreting $(a,b,c,d)$ as the creation operators for $(a,b,c,d)$ preons and by interpreting $\mathcal{D}^j_{mm'}$ as the creation operator for a composite particle that is a superposition of substructures with varying numbers of preons according to the following four equations$^{(7)}$

\be
\left( t,t_3,t_0,Q \right) = \sum_{p=(a,b,c,d)} n_p \left(t_p,t_{3p},t_{0p},Q_p \right)
\ee

\noindent where the $(t,t_3,t_0,Q)$ are the isotopic spin and charge indices of the composite particle $\mathcal{D}^j_{mm'}$, and the $(t_p,t_{3p},t_{0p},Q_p)$ are the corresponding indices of the $p^{th}$ preon, all determined by (5.1).  Here $n_p$ is the exponent in (5.3) of the creation operator for the $p^{th}$ preon and hence records the number of $p$ preons in the different substructures appearing in (5.3).

\hspace{10mm}  The relative weights of the various subamplitudes depend on $\alpha$ through $\mathcal{A}^j_{mm'} \left(\alpha,s,t \right)$ in the expansion (5.3).

\section{Eigenstates of Energy}

\hspace{10mm} There is also a dynamical interpretation of $\alpha$ as it appears in the difference operator $D_t$, defined by

\be
\mathcal{D}_t \Psi \left (t \right) = \frac{\Psi \left( \alpha t \right) - \Psi \left( t \right)}{\alpha t - t}
\ee

\noindent $D_t$ satisfies the $SLq(2)$ invariant commutator$^{(7)}$

\be
\mathcal{D}_t t - \alpha t \mathcal{D}_t = 1
\ee

\noindent In the limit $\alpha \rightarrow 1$

\be
\mathcal{D}_t \rightarrow \frac{\partial}{\partial t}
\ee

\noindent and

\be
P_t \equiv \frac{\hbar}{i} \mathcal{D}_t \rightarrow \frac{\hbar}{i} \frac{\partial}{\partial t}
\ee

\noindent  Then $P_t$ becomes the usual energy operator, where $t$ is time, while $-i\hbar$ (6.2) becomes the usual Heisenberg commutator.

\hspace{10mm}  The eigenstates of the energy operator $P_t$ are the following twisted exponentials

\be
\mathcal{E}_\alpha \left( i \omega t \right) = \sum \frac{(i \omega t)^n}{\langle n \rangle_\alpha !}
\ee

\noindent with

\be
\langle n \rangle_\alpha = \frac{\alpha^n -1}{\alpha - 1}
\ee

\noindent by (6.1), (6.5), and (6.6), i.e.

\be
P_t \mathcal{E}_\alpha \left(i \omega t \right) = \hbar \omega \mathcal{E}_\alpha \left( i \omega t \right)
\ee

\hspace{10mm} The eigenvalue of the energy operator, $P_t$, is $E=\hbar \omega$ but the energy eigenfunction is no longer a simple harmonic function with frequency $\frac{E}{\hbar}$.  The time-dependence of the quantum state is instead given by (6.5), again an expression of $SLq(2)$ symmetry, and the parameter $\alpha$ appearing in (6.5) may now be understood as $\alpha = \frac{e}{g}$.  In this way, the time-dependence of an atomic clock would be influenced by the ratio of the weak and strong couplings.

\ve
\section{References}

\begin{enumerate}
\item R. J. Finkelstein Int. J. Mod. Phys. A $\underline{20}$ 487 (2005)
\item A. C. Cadavid and R.J. Finkelstein Int. J. Mod. Phys. A $\underline{21}$ (2006)
\item R. J. Finkelstein arXiv: 1011.2545 v1 hep-th
\item R. J. Finkelstein Int. J. Mod. Physics A $\underline{24}$ (2007)
\item R. J. Finkelstein Int. J. Mod. Physics A $\underline{24}$ (2009)
\item R. J. Finkelstein arXiv: 0901.1687 hep-th
\item R. J. Finkelstein arXiv: 0912.3552 v2 hep-th
\item Julian Schwinger Science $\underline{165}$ 757-761 (1969)
\item R. J. Finkelstein arXiv: 1011.076 v1 hep-th

\end{enumerate}

\end{document}